\documentclass[twocolumn,superscriptaddress, showpacs] {revtex4}
\usepackage{graphicx}
\usepackage{amsmath}

\begin{document}

\title[Short Title]{One-step implementation of the genuine Fredkin gate  in high-$Q$ coupled three-cavity arrays}
\author{Xiao-Qiang Shao\footnote{E-mail: xqshao@yahoo.com}}
\affiliation{School of Physics, Northeast Normal University
Changchun 130024, People's Republic of China}
\affiliation{Centre for Quantum Technologies, National University of Singapore, 3 Science Drive 2, Singapore 117543}
\author{Tai-Yu Zheng}
\affiliation{School of Physics, Northeast Normal University
Changchun 130024, People's Republic of China}
\author{Xun-Li Feng}
\affiliation{Centre for Quantum Technologies, National University of Singapore, 3 Science Drive 2, Singapore 117543}
\author{C. H. Oh}
\affiliation{Centre for Quantum Technologies, National University of Singapore, 3 Science Drive 2, Singapore 117543}
\author{Shou Zhang}
 \affiliation{Department of Physics, College of Science,
Yanbian University, Yanji, Jilin 133002, People's Republic of China}

\begin{abstract}
We present two efficient methods for implementing the Fredkin gate  with atoms separately trapped in an array of three high-$Q$ coupled cavities.
The first proposal is based on the resonant dynamics, which leads to a fast resonant interaction in a certain subspace while leaving others unchanged, and the second one utilizes a dispersive interaction such that the effective long-distance dipole-dipole interaction between two distributed target
qubits is achieved by virtually excited process. Both schemes can achieve the standard form of the Fredkin gate in a single step without any subsequent single-qubit operation.  The effects of decoherence on the performance of the gate are also analyzed in virtue of master equation, and the strictly numerical simulation reveals that
 the average fidelity of the quantum gate is high.
\end{abstract}
 \maketitle \maketitle

\section{introduction}
The quantum mechanics based computers  can  outperform  traditional computers by a far greater order of magnitude in computing power, because they permit parallel computation due to the principle of superposition and entanglement \cite{shor1,shor,Grover}.
In a quantum computer, the quantum logic gates  constitute the basic building blocks of quantum circuit which are reversible transformations on an $n$-qubit register.
According to the principle of universal quantum computation, any unitary operation can be decomposed into a series of single qubit operations along with two-qubit gates \cite{DPD}, e.g. achievement of a generic two-qubit gate may require a sequence
of up to three CNOT gates combined with single-qubit
rotations. Nevertheless, this kind of decomposition becomes inefficient as applied to a multi-qubit (three or more) gate, for the procedure will become more complicated and makes the quantum system further susceptible to the environment. Therefore, much attention has been paid on the direct implementation of multi-qubit gates both in theory and experiment \cite{yangcp,yangcp1,xzg1,gwl,shao,shao2,wly,yang,yang1,xzg,am,cj}.
\begin{eqnarray}\label{fred}
{U_{\rm FRED}}=\left[\begin{array}{c c c c c c c c}
1 & 0 & 0 & 0 & 0 & 0 & 0 & 0\\
0 & 1 & 0 & 0 & 0 & 0 & 0 & 0\\
0 & 0 &1 & 0 & 0 & 0 & 0 & 0\\
0 & 0 & 0 & 1 & 0 & 0 & 0 & 0\\
0 & 0 & 0 & 0 & 1 & 0 & 0 & 0\\
0 & 0 & 0 & 0 & 0 & 0 & 1 & 0 \\
0 & 0 & 0 & 0 & 0 & 1& 0 & 0 \\
0 & 0 & 0 & 0 & 0 & 0 & 0 & 1 \\
\end{array}
\right].
\end{eqnarray}

The Fredkin gate \cite{fredkin}, together with Toffoli gate becomes the important logic gate in the domain of three-qubit gates. Its matrix form expanded in subspace $\{|0_2\rangle,|1_2\rangle,|0_1\rangle,|1_1\rangle,|0_3\rangle,|1_3\rangle\}$ is shown in Eq.~(\ref{fred}), from which we see two target qubits  swap their information $|01\rangle_{1,3}\Leftrightarrow|10\rangle_{1,3}$ if and only if the control qubit is in $|1_2\rangle$.
  This gate not only has been useful in designing a circuit for error correcting quantum computations \cite{barenco}, but also has given a
simple implementation of the quantum computer to solve
Deutsch's problem \cite{cy}. Since the first quantum optical Fredkin gate was proposed in theory by Milburn \cite{mil}, the physical realization of Fredkin gate has flourished in the field of linear optics in recent years. Fiur\'{a}\v{s}ek suggested a heralded Fredkin gate that used ancilla photons, interference, and single-photon detection to emulate the the cross-Kerr nonlinearity \cite{JF}, then the author put forward  an alternative scheme for linear optical quantum Fredkin gate based on the combination of recently experimentally demonstrated linear optical partial-SWAP gate and controlled-Z gates \cite{JF1}.  Gong {\it et al.} presented two methods for a linear optical quantum Fredkin gate  with only linear optics and single photons \cite{yxgong}. Lin {\it et al.} realized the Fredkin gate
with weak cross-Kerr nonlinearity based on the controlled-path gate \cite{Linq}. The prominent merit of linear-optical technique is that  photons is best for the quantum information process due to its weak interaction with the environment, but the low success probability may restraint the development of large-scale quatum computation. It is worth noting that Ref.~\cite{Linq1} presented a simple architecture for deterministic quantum circuits operating on single photon
qubits.
The research about the deterministic Fredkin gate initiated in several hybrid quantum systems, e.g.  ion-phonon in a two-dimensional ion trap \cite{zou} and  atom-photon via the cavity input-output process \cite{wang,jsong}. However, the different representations of qubits do not agree with the requirement for quantum computing as a typical quantum algorithm may require each qubit to be treated equally.  Although there are certain schemes that simulate Fredkin gate operation in cavity QED system \cite{zheng} and superconduction system \cite{shao1}, either extra ancillary levels need to be introduced or local operations should be applied.

\begin{figure}
\scalebox{0.4}{\includegraphics{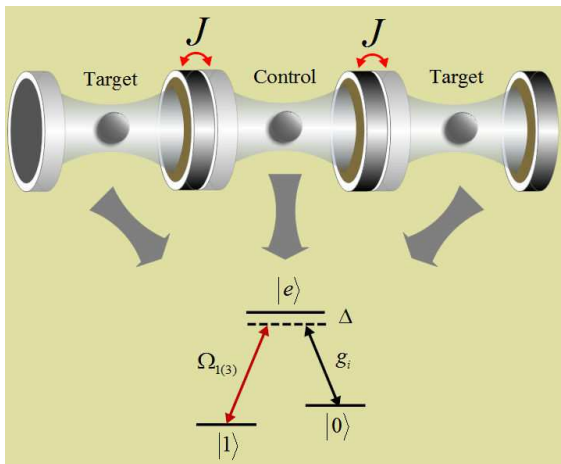}} \caption{\label{p0}
(Color online) Schematic of three atoms trapped in three coupled cavities.
Each qubit is encoded into two lower-energy levels labeled as $|0\rangle$ and $|1\rangle$. The transitions between the levels
$|e_i\rangle\leftrightarrow|0_i\rangle$ is coupled to the cavity
mode with the coupling constants $g_i$, and the
transitions $|e_{1(3)}\rangle\leftrightarrow|1_{1(3)}\rangle$
 is driven by
a classical pulse with the Rabi frequencies
$\Omega_{1(3)}$, $\Delta$ represents the
corresponding one-photon detuning parameter  and the photon can hop between two cavities with coupling strength $J$. The atom in the middle cavity serves as the control qubit while the atoms in bilateral cavities play the role of the target qubits, and we label the atoms as 1, 2, 3 from left to right for convenience, and the state $|\alpha_1,\beta_2,\gamma_3\rangle_a$ is abbreviated  to $|\alpha,\beta,\gamma\rangle_a$ in the text. }
\end{figure}

Coupled-cavity models describe a series of optical cavities, each containing one or more atoms with photons permitted to hop between neighboring cavities. The original purpose for introducing coupled-cavity models is to overcome the problem of individual addressability. Along with theoretically intensive study on this model, many interesting phenomena are observed such as photon blockade induced Mott transitions \cite{bose}, polaritonic characteristics of insulator and superfluid states \cite{Kim}, fractional quantum Hall state \cite{bose2}, multimode entanglement \cite{liew}, and other quantum many-body phenomena \cite{mb}. Recently, the coupled-cavity systems are also applied in distributed controlled-phase gate \cite{asera,zqyin,zbyang,zbyang1}.
In this paper, we explore the particular feature in coupled three-cavity array system and propose two efficient schemes for implementing the genuine Fredkin gate in a coupled-cavity array. Compared with previous proposals, the advantages of ours are
threefold: (i) the quantum information is encoded into three identical atoms without introducing any ancillary level, which satisfies the requirement for quantum computation; (ii) the qubits are trapped separately in three cavities, and this arrangement will make it convenient to control and measure
qubit individually; (iii) the Fredkin gate is fast achieved for it only needs one-step operation dispense with any single qubit gate. In particular, the present scheme provide
two options for achieving high-fidelity quantum gate without resorting to non-identical coupling adopted in Ref~\cite{shao1}.

The structure of the paper is as follows. In Sec.~2, we derive the effective Hamiltonians that govern the evolution of quantum states for the
resonant- and dispersive interaction between atom and coupled cavity, respectively. In Sec.~3, we discuss the performance of the scheme via the definition of average fidelity for quantum logic gate and consider the effect of typical decoherence in virtue of master equation. Then a summary appears in Sec.~4.

\section{Effective Hamiltonian in coupled-cavity array system}

The considered physical system consists of three atoms with $\Lambda$-type configuration
trapped in a coupled-cavity array, as shown in Fig.~\ref{p0}.
Each atom interacts with cavity mode via the Jaynes-Cummings (JC) model, where
the transitions between the levels
$|e_i\rangle\leftrightarrow|0_i\rangle$ are coupled to the cavity
mode with the coupling constants $g_i$. In addition, we apply two classical fields to drive the
atomic transitions $|e_{1}\rangle\leftrightarrow|1_{1}\rangle$ and $|e_{3}\rangle\leftrightarrow|1_{3}\rangle$, respectively.
 The
corresponding one-photon detuning parameter is $\Delta$ and the photon can hop between two neighbor cavities with coupling strength $J$.
The Hamiltonian of the system in the Schr\"{o}diner picture reads ($\hbar = 1$)
\begin{eqnarray}\label{initial01}
{H}_S&=&
\sum_{j=1,3}\Omega_j(|e_j\rangle\langle1_j|e^{-i\omega_l^jt}+|1_j\rangle\langle e_j|e^{i\omega_l^jt})
\nonumber\\&&+\sum_{k=1}^{2}J(a_k^{\dag}a_{k+1}+a_ka^{\dag}_{k+1})\nonumber\\&&
+\sum_{i=1}^{3}g_i(a_i|e_i\rangle\langle
0_i|+|0_i\rangle\langle e_i|a_i^{\dag})+\sum_{k=1}^3\omega_ca_k^{\dag}a_k\nonumber\\&&+\sum_{i=1}^3\omega_e|e_i\rangle\langle e_i|+\omega_0|0_i\rangle\langle 0_i|
+\omega_1|1_i\rangle\langle 1_i|,
\end{eqnarray}
where $\omega_l^j$ is the frequency of  $j$th classical field, $\omega_c$ denotes the cavity frequency and $\omega_{e(0,1)}$ represents the energy of atomic level $|e\rangle(|0\rangle,|1\rangle)$.
In the interaction
picture, after performing a rotating with respect to $U=\exp(i\Delta t\sum_{i=1}^3|e_i\rangle\langle e_i|)$, the Hamiltonian can be written as
\begin{eqnarray}\label{initial0}
{H}_I&=&
\sum_{j=1,3}\Omega_j(|e_j\rangle\langle1_j|+|1_j\rangle\langle e_j|)
+\sum_{k=1}^{2}J(a_k^{\dag}a_{k+1}\nonumber\\&&+a_ka^{\dag}_{k+1})
+\sum_{i=1}^{3}g_i(a_i|e_i\rangle\langle
0_i|+|0_i\rangle\langle e_i|a_i^{\dag})\nonumber\\&&+\Delta|e_i\rangle\langle e_i|.
\end{eqnarray}
The last term in Eq.~(\ref{initial0}) plays an important role in our scheme because the presence of $\Delta$ or not determines the dynamics to be  resonant or dispersive. In what follows, we will discuss the possibility for one-step achieving the genuine Fredkin with the mentioned dynamics in detail.
\subsection{Resonant interaction between cavity and doped atom}
In this section, we focus on synthesizing the Fredkin gate with the resonant interaction. The system we consider is a simplification of Fig.~\ref{p0}  ($\Delta=0$), i.e. the atom in each cavity is resonantly coupled to the cavity field and the classical field. Before preceding,
we first divide the Hamiltonian of Eq.~(\ref{initial0}) into two parts as analogous to quantum Zeno dynamics \cite{Paolo1}:
\begin{eqnarray}\label{200}
H_I&=&H_1+H_2,\nonumber\\
H_1&=&\sum_{j=1,3}\Omega_j(|e_j\rangle\langle1_j|+|1_j\rangle\langle e_j|),\nonumber\\
H_2&=&\sum_{k=1}^{2}J(a_k^{\dag}a_{k+1}+a_ka^{\dag}_{k+1})+\sum_{i=1}^{3}g_i(a_i|e_i\rangle\langle
0_i|\nonumber\\&&+|0_i\rangle\langle e_i|a_i^{\dag}),
\end{eqnarray}
where $H_1$ is the Hamiltonian of the system to be investigated and
$H_2$ can be considered as an additional interaction Hamiltonian
performing the ``measurement". For a large ratio $g/\Omega_{1(3)}(J/\Omega_{1(3)})$,
the system investigated can be viewed as
dominated by the evolution operator
$
{\cal{U}}(t)=\exp(iH_2t)U_I^{re}(t),
$
which can be shown to have the form
$
{\cal{U}}(t)=\exp(-iH_Zt),
$
where
$
H_Z=\sum_nP_nH_1P_n,
$
is called Zeno Hamiltonian, $P_n$ being the eigenprojection of
$H_2$ belonging to the eigenvalue $\eta_n$, and $
H_2=\sum_n\eta_nP_n
$, thus the total Hamiltonian reads
\begin{equation}\label{use}
H^{re}_{eff}=\sum_n\eta_nP_n+P_nH_1P_n.
\end{equation}
If the interested quantum system is freezed to the dark subspace $(n=0)$ of $H_2$, the whole system are governed by the effective Hamiltonian
\begin{equation}\label{use1}
H^{re}_{eff}=P_0H_1P_0.
\end{equation}

For three-qubit logic gate, there are eight input states needed to be computed  in subspace $\{|0_1\rangle,|1_1\rangle,|0_2\rangle,|1_2\rangle,|0_3\rangle,|1_3\rangle\}$. Since the JC model Hamiltonian of atom and coupled cavity conserves the  excitation number ${\cal N}_k=a_{k}^{\dag}a_k+|e_k\rangle\langle e_k|$, it is convenient to reclassify the above bases in different excitation subspaces, i.e. $\{|000\rangle_a|000\rangle_c,|010\rangle_a|000\rangle_c\}$ belong to the ``zero excitation" subspace,
$\{|001\rangle_a|000\rangle_c, |100\rangle_a|000\rangle_c\}$ and $\{|011\rangle_a|000\rangle_c, |110\rangle_a|000\rangle_c\}$ correspond to two independent ``single excitation" subspaces, and $\{|101\rangle_a|000\rangle_c\}$ and $\{|111\rangle_a|000\rangle_c\}$ are included in two different ``two excitation" subspaces.  The primary function of a Fredkin gate is to perform the controlled swap operation
$|1\rangle_C|01\rangle_T\Leftrightarrow|1\rangle_C|10\rangle_T$, thereby the dynamical evolutions of $\{|011\rangle_a|000\rangle_c, |110\rangle_a|000\rangle_c\}$ are investigated first.

In the closed subspace $\{|011\rangle_a|000\rangle_c,$
$|01e\rangle_a|000\rangle_c,$
$|010\rangle_a|001\rangle_c,$
$ |010\rangle_a|010\rangle_c,$
$|010\rangle_a|100\rangle_c,$
$|e10\rangle_a|000\rangle_c,$
$|110\rangle_a|000\rangle_c\}$, we can expand Hamiltonian of Eq.~(\ref{initial0}) as
\begin{eqnarray}\label{066}
M^{re}=\left[\begin{array}{c c c c c c c}
0 & \Omega_3 & 0 & 0 & 0 & 0 & 0\\
\Omega_3 & 0 & g & 0 & 0 & 0 & 0\\
0 & g & 0 & J & 0 & 0 & 0\\
0 & 0 & J & 0 & J & 0 & 0\\
0 & 0 & 0 & J & 0 & g & 0\\
0 & 0 & 0 & 0 & g & 0 & \Omega_1\\
0 & 0 & 0 & 0 & 0 & \Omega_1 & 0\\
\end{array}
\right].
\end{eqnarray}

The eigenprojection operator $P_n$ of atom-cavity interaction can be constructed from
$
|E_1\rangle=\frac{1}{\sqrt{3}}\big(|e10\rangle_a|000\rangle_c+|01e\rangle_a|000\rangle_c-|010\rangle_a|010\rangle_c\big),
$
$
|E_2\rangle=\frac{1}{2}\big[\big(|e10\rangle_a-|01e\rangle_a)\big|000\rangle_c-|010\rangle_a\big(|100\rangle_c-|001\rangle_c\big)\big],
$
$
|E_3\rangle=\frac{1}{2}\big[\big(|e10\rangle_a-|01e\rangle_a)\big|000\rangle_c+|010\rangle_a\big(|100\rangle_c-|001\rangle_c\big)\big],
$
$
|E_4\rangle=-\frac{1}{\sqrt{12}}\big(|e10\rangle_a+|01e\rangle_a)\big|000\rangle_c+\frac{1}{2\sqrt{3}}|010\rangle_a\big(\sqrt{3}|100\rangle_c+\sqrt{3}|001\rangle_c-2|010\rangle_c\big),
$
and $
|E_5\rangle=\frac{1}{\sqrt{12}}\big(|e10\rangle_a+|01e\rangle_a)\big|000\rangle_c+\frac{1}{2\sqrt{3}}|010\rangle_a\big(\sqrt{3}|100\rangle_c+\sqrt{3}|001\rangle_c+2|010\rangle_c\big)
$
corresponding to the eigenvalues
$E_1=0,$ $E_2=-g,$ $E_3=g,$ $E_4=-\sqrt{3}g,$ and $E_5=\sqrt{3}g,$ where we have assumed $J=g$ for simplicity. Thus the Eq.~(\ref{066}) can be rewritten as
\begin{eqnarray}\label{initialt}
{H}_{re}&=&\Omega_1|110\rangle_a|000\rangle_c\bigg[\frac{1}{\sqrt{3}}\langle E_1|+\frac{1}{2}\big(\langle E_2|+\langle E_3|\big)\nonumber\\&&-\frac{1}{\sqrt{12}}\big(\langle E_4|-\langle E_5|\big)\bigg]+\Omega_3|011\rangle_a|000\rangle_c\bigg[\frac{1}{\sqrt{3}}\langle E_1|\nonumber\\&&-\frac{1}{2}\big(\langle E_2|+\langle E_3|\big)-\frac{1}{\sqrt{12}}\big(\langle E_4|-\langle E_5|\big)\bigg]\nonumber\\&&+{\rm H.c.}+\sum_{i=1}^5E_i|E_i\rangle\langle E_i|.
\end{eqnarray}
Eq.~(\ref{initialt}) describes that  $|110\rangle_a|000\rangle_c$ and $|011\rangle_a|000\rangle_c$ couple to excitation state $|E_i\rangle$ with detuning $E_i$, respectively. Thus under a strong continuous coupling $g\gg|\Omega_{1(3)}|$, we may neglect the large detuned terms and the dominant part that governs the evolution of quantum states in this subspace reduces to the resonant coupling with detuning $E_1=0$, i.e.
\begin{equation}\label{eeff}
H^{re}_{eff}=\frac{1}{\sqrt{3}}\big(\Omega_1|110\rangle_a+\Omega_3|011\rangle_a\big)|000\rangle_c\langle D|+{\rm H.c.},
\end{equation}
where we have relabeled $|E_1\rangle$ as $|D\rangle$ for it is immune to the interaction between atoms and coupled-cavity array.

\begin{figure}
\scalebox{0.5}{\includegraphics{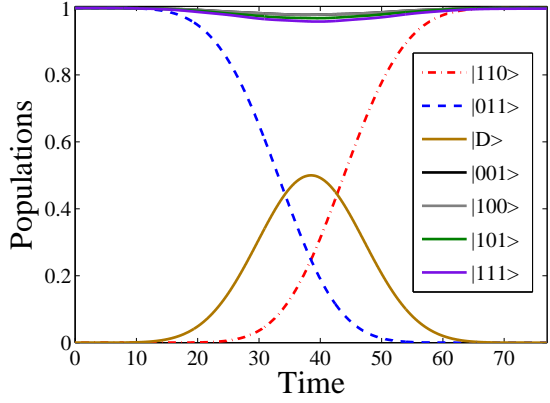}} \caption{\label{p5}(Color online) Evolutions for the populations of interested states.
The qubit states $|011\rangle_a|000\rangle_c$ and $|110\rangle_a|000\rangle_c$ coherently transform into each other via the media state $|D\rangle$, while other qubit states are freezed to their original status due to the large detuned interaction. The corresponding parameters are $J=g$ and $\Omega_{max}=0.05g$.  }
\end{figure}

According to Eq.~(\ref{use1}), we find other input states  do not participate in the dynamical evolution at the requirement of Zeno condition $|\Omega_{1(3)}|\ll g$.
Therefore a genuine Fredkin gate can be carried out in one step as
\begin{equation}\label{eefff}
\int_0^T\frac{1}{\sqrt{3}}\Omega {\rm d}t=\frac{\pi}{\sqrt{2}}, \  \  \  \  (\Omega_1=-\Omega_3=\Omega).
\end{equation}
We may control two Rabi frequencies of classical fields adiabatically as \cite{beige}
\begin{equation}\label{use}
\Omega(t)=2\Omega_{max}\sin^2\bigg[\sqrt{\frac{2}{3}}\Omega_{max}t\bigg],
\end{equation}
and this choice corresponds to the interaction time $T=\sqrt{3}\pi/(\sqrt{2}\Omega_{max})$. In Fig.~\ref{p5}, we depict
the time evolution of each input state during the gate operation process with the full Hamiltonian of Eq.~(\ref{initial0}). We find the Zeno condition $|\Omega_{1(3)}|\ll g$ guarantees the effectiveness of Eq.~(\ref{eeff}) well and  an adiabatic evolution of quantum gate makes the scheme more stable, because the populations of states $|011\rangle_a|000\rangle_c$, and $|110\rangle_a|000\rangle_c$ keep maximal in a period of time instead of one spot. The prominent advantage of the present scheme is that the Fredkin gate is achieved in a short time due to the resonant interaction between coupled-cavity array and doped atom, but the dark state $|D\rangle$ still incorporates the excited state of atoms and the middle cavity, which may cause the quantum gate to be sensitive to the decoherence. In next section, we will modify the above scheme and resort to another method for implementing the Fredkin gate in a dispersive way.

\subsection{Dispersive interaction between cavity and doped atom}
In this section, we discuss the possibility of achieving the Fredkin gate via dispersive interaction between  atom and coupled cavity. The analysis still relies on the idea of quantum Zeno dynamics, nevertheless, more input states should be considered since the complicated dynamics induced by $\Delta$.
For the subspace $\{|100\rangle_a|000\rangle_c$,
$|001\rangle_a|000\rangle_c$,
$|000\rangle_a|100\rangle_c$,
$|000\rangle_a|010\rangle_c,$
$|000\rangle_a|001\rangle_c$,
$|e00\rangle_a|000\rangle_c$,
$|0e0\rangle_a|000\rangle_c$,
$|00e\rangle_a|000\rangle_c\}$, we need to introduce the following collective cavity states
$
|\varphi_a\rangle=\frac{1}{2}|000\rangle_a(|100\rangle-{\sqrt{2}}|010\rangle+|001\rangle)_c,
|\varphi_b\rangle=\frac{1}{2}|000\rangle_a(|100\rangle+{\sqrt{2}}|010\rangle+|001\rangle)_c,
|\varphi_c\rangle=\frac{1}{\sqrt{2}}|000\rangle_a(|100\rangle-|001\rangle)_c,
$
and collective atomic states
$
|\varphi_1\rangle=\frac{1}{2}(|e00\rangle-{\sqrt{2}}|0e0\rangle+|00e\rangle)_a|000\rangle_c,
|\varphi_2\rangle=\frac{1}{2}(|e00\rangle+{\sqrt{2}}|0e0\rangle+|00e\rangle)_a|000\rangle_c,
|\varphi_3\rangle=\frac{1}{\sqrt{2}}(|e00\rangle-|00e\rangle)_a|000\rangle_c.
$
In the new basis $\{|\varphi_1\rangle$,
$|\varphi_a\rangle$,
$|\varphi_2\rangle$,
$|\varphi_b\rangle$,
$|\varphi_3\rangle$,
$|\varphi_c\rangle\},$ the Hamiltonian of atom-cavity interaction is simplified to the form of three $2\times2$ block matrices:
\begin{eqnarray}\label{begin}
M^{de}=\left[\begin{array}{c c c c c c}
\Delta & g & 0 & 0 & 0 & 0 \\
g & -\sqrt{2}J & 0 & 0 & 0 & 0\\
0 & 0 & \Delta & g & 0 & 0\\
0 & 0 & g & \sqrt{2}J & 0 & 0\\
0 & 0 & 0 & 0 & \Delta & g \\
0 & 0 & 0 & 0 & g & 0 \\
\end{array}
\right],
\end{eqnarray}
and this matrix can be solved easily of which the  eigenvalues are
\begin{eqnarray}\label{g}
&&E_1=\frac{1}{2}\left(\Delta-\sqrt{2}J-\sqrt{4g^2+\Delta^2+2\sqrt{2}\Delta J+2J^2}\right),\nonumber\\
&&E_2=\frac{1}{2}\left(\Delta-\sqrt{2}J+\sqrt{4g^2+\Delta^2+2\sqrt{2}\Delta J+2J^2}\right),\nonumber\\
&&E_3=\frac{1}{2}\left(\Delta+\sqrt{2}J-\sqrt{4g^2+\Delta^2-2\sqrt{2}\Delta J+2J^2}\right),\nonumber\\
&&E_4=\frac{1}{2}\left(\Delta+\sqrt{2}J+\sqrt{4g^2+\Delta^2-2\sqrt{2}\Delta J+2J^2}\right),\nonumber\\
&&E_5=\frac{1}{2}\left(\Delta-\sqrt{4g^2+\Delta^2}\right),\nonumber\\
&&E_6=\frac{1}{2}\left(\Delta+\sqrt{4g^2+\Delta^2}\right),
\end{eqnarray}
corresponding to the eigenstates
\begin{eqnarray}\label{g}
&&|E_1\rangle=\frac{\alpha}{\sqrt{4g^2+\alpha^2}}|\varphi_1\rangle+\frac{2g}{\sqrt{4g^2+\alpha^2}}|\varphi_a\rangle,\nonumber\\
&&|E_2\rangle=\frac{2g}{\sqrt{4g^2+\alpha^2}}|\varphi_1\rangle-\frac{\alpha}{\sqrt{4g^2+\alpha^2}}|\varphi_a\rangle,\nonumber\\
&&|E_3\rangle=\frac{\beta}{\sqrt{4g^2+\beta^2}}|\varphi_2\rangle+\frac{2g}{\sqrt{4g^2+\beta^2}}|\varphi_b\rangle,\nonumber\\
&&|E_4\rangle=\frac{2g}{\sqrt{4g^2+\beta^2}}|\varphi_2\rangle-\frac{\beta}{\sqrt{4g^2+\beta^2}}|\varphi_b\rangle,\nonumber\\
&&|E_5\rangle=-\sqrt{\frac{\sqrt{4g^2+\Delta^2}-\Delta}{2\sqrt{4g^2+\Delta^2}}}|\varphi_3\rangle+\sqrt{\frac{\sqrt{4g^2+\Delta^2}+\Delta}{2\sqrt{4g^2+\Delta^2}}}|\varphi_c\rangle,\nonumber\\
&&|E_6\rangle=\sqrt{\frac{\sqrt{4g^2+\Delta^2}+\Delta}{2\sqrt{4g^2+\Delta^2}}}|\varphi_3\rangle+\sqrt{\frac{\sqrt{4g^2+\Delta^2}-\Delta}{2\sqrt{4g^2+\Delta^2}}}|\varphi_c\rangle,
\end{eqnarray}
where the coefficients are
\begin{eqnarray}\label{g}
&&\alpha=\Delta+\sqrt{2}J-\sqrt{4g^2+\Delta^2+2\sqrt{2}\Delta J+2J^2},\nonumber\\
&&\beta=\Delta-\sqrt{2}J-\sqrt{4g^2+\Delta^2-2\sqrt{2}\Delta J+2J^2}.
\end{eqnarray}
Clearly, the qubit states $|100\rangle_a|000\rangle_c$ and $|001\rangle_a|000\rangle_c$ couple to the above eigenstates through two classical fields $\Omega_1$ and $\Omega_3$.
If the large detuning condition is satisfied, i.e. $\{|E_i|, |E_i-E_j|\}\gg\{|\Omega_1|,|\Omega_3|\}$, the effective coupling strength between $|100\rangle_a|000\rangle_c$ and $|001\rangle_a|000\rangle_c$ is calculated by summing six independent-eigenstate transition channels.
Most interestingly, if the one-photon detuning parameter $\Delta$, the coupling strength between cavities $J$ and the atom-cavity interacting constant meet the condition $\Delta=J=g$, a concise form of dipole-dipole interaction is obtained as
\begin{eqnarray}\label{qw}
H^1_{eff}&=&\frac{\Omega^2_1}{g}|100\rangle\langle100|+\frac{\Omega^2_3}{g}|001\rangle\langle001|\nonumber\\&&+\frac{\Omega_1\Omega_3}{g}\big(|100\rangle\langle001|
+{\rm H.c.}\big).
\end{eqnarray}
Likewise, the evolution of quantum states $|110\rangle_a|000\rangle_c$ and $|011\rangle_a|000\rangle_c$ are governed by
\begin{eqnarray}\label{wq}
H^2_{eff}&=&-\frac{\Omega^2_1}{2g}|110\rangle\langle110|-\frac{\Omega^2_3}{2g}|011\rangle\langle011|\nonumber\\&&-\frac{\Omega_1\Omega_3}{2g}\big(|110\rangle\langle011|
+{\rm H.c.}\big).
\end{eqnarray}
Eqs.~(\ref{qw}) and (\ref{wq}) describe two kind of dipole-dipole interaction with different coupling constants. A permutation
 between states $|110\rangle_a|000\rangle_c$ and $|011\rangle_a|000\rangle_c$ is exactly followed by a cycle oscillation of state $|100\rangle_a|000\rangle_c$ or $|001\rangle_a|000\rangle_c$.
The dynamical behaviors of states $|101\rangle_a|000\rangle_c$ and $|111\rangle_a|000\rangle_c$ are derived in the same way, but we do not need to consider their effects in the final effective Hamiltonian because the stark-shifts for them are cancelled due to the
destructive intereference in respective subspaces. Thus a selection of interaction time $T={g\pi}/{\Omega^2}$ $(\Omega_1=-\Omega_3=\Omega)$ will lead to the stand form of the Fredkin gate.

\begin{figure}
\scalebox{0.4}{\includegraphics{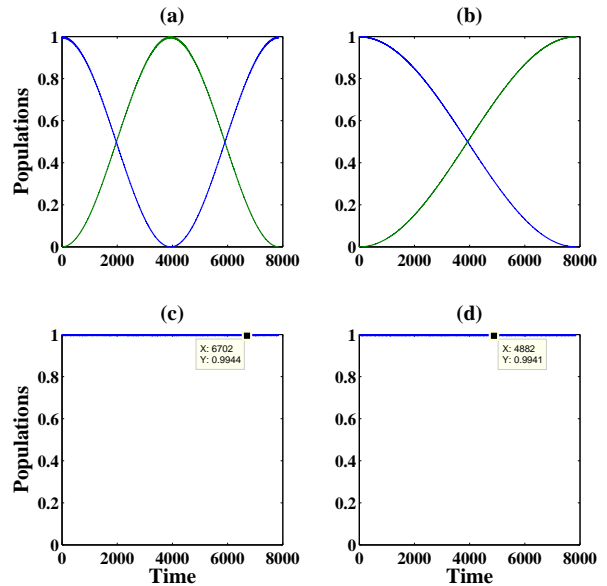}} \caption{\label{p3}
(Color online) Evolution of the system from exact calculations of Hamiltonian Eq.~(\ref{initial0}).
(a) these two curves represent the conversion of populations between states $|001\rangle$ and $|100\rangle$;
(b) the populations of states $|011\rangle$ and $|110\rangle$;
(c) the population of state $|101\rangle$;
(d) the population of state $|111\rangle$. All cavities stay in vacuum states and the data listed in the data cursor further validate the effectiveness of our approximation.
 The revelant parameters are set as $\Omega_1=-\Omega_3=\Omega=0.02g$, and $\Delta=J=g$. The dimensionless time is measured in the unit of $g^{-1}$.}
\end{figure}

To illustrate the effectiveness of our approximation, we plot the population of each qubit state during the evolution in Fig.~\ref{p3} with the full Hamiltonian (\ref{initial0}), where the subplot (a) describes the conversion between states $|001\rangle_a|000\rangle_c$ and $|100\rangle_a|000\rangle_c$, (b) represents
$|011\rangle_a|000\rangle_c$ and $|110\rangle_a|000\rangle_c$, and (c) and (d) depict the populations for states $|101\rangle_a|000\rangle_c$ and $|111\rangle_a$$|000\rangle_c$, respectively. It is unequivocal that the agreement between the
effective and exact model is excellent under the given parameters and the data listed in the data cursor further validate the effectiveness of our approximation.

\section{Discussion and experimental feasibility}
\begin{figure}
\scalebox{0.32}{\includegraphics{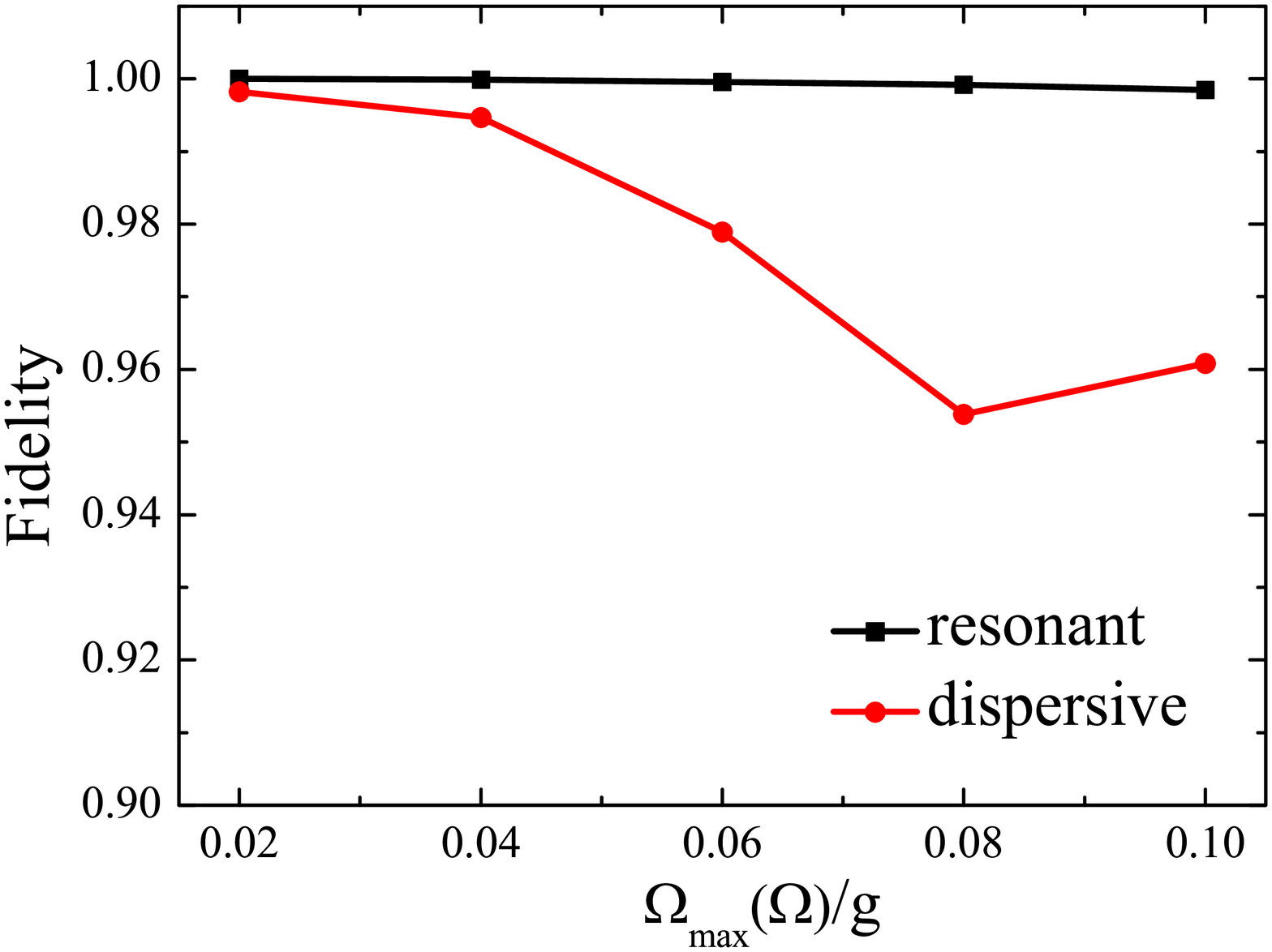}}\caption{\label{px}
(Color online) The average gate fidelity versus the Zeno requirement $\Omega_{max}(\Omega)/g$. The black curve represents the resonant model and the red
curve corresponds to the dispersive model.}
\end{figure}

The average gate fidelity is introduced to qualify the performance of the Fredkin gate \cite{Measure,Mea1}
\begin{eqnarray}\label{He}
\overline{F}(\varepsilon,U_{\rm FRED})=\frac{\sum_j{\rm tr}\big[U_{\rm FRED}U_j^\dag
U_{\rm FRED}^\dag \varepsilon(U_j)\big]+d^2}{d^2(d+1)},
\end{eqnarray}
where $d=8$ for three qubits and $U_j$ being the tensor of Pauli
matrices  $III, IIX,IIY,\cdots, ZZZ$, $U_{\rm FRED}$ being the ideal Fredkin
gate and $\varepsilon$ being the trace-preserving quantum
operation obtained through our Fredkin gate. In Fig.~\ref{px}, we plot the relation between average gate fidelities for both resonant model and dispersive model versus the ratio $\Omega_{max}(\Omega)/g$, respectively. In the range of $0.02\leq\Omega_{max}(\Omega)/g\leq0.1$, the fidelity of resonant Fredkin gate keeps value much higher than 99\%, while the fidelity for dispersive Fredkin gate decreases
from 99.82\% to 96\% in an oscillation form.
\begin{figure}
\scalebox{0.32}{\includegraphics{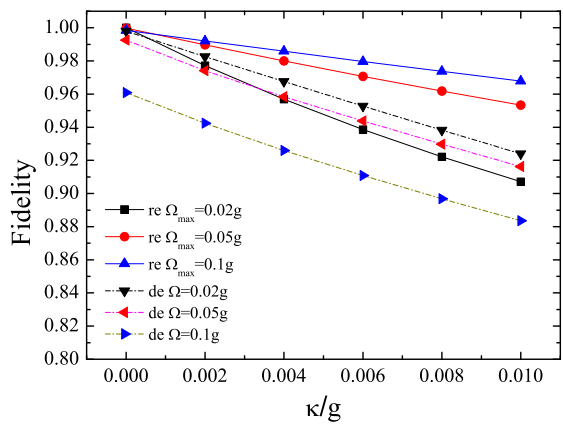}}\caption{\label{pw}
(Color online) The average gate fidelities for resonant model and dispersive model versus decoherence parameter $\kappa/g$ under three different driving lasers respectively, where we have set $\gamma=\kappa$ and $\Delta=J=g$. }
\end{figure}
Now we pay attention to
the decoherence effects on our Fredkin operation. In the model of coupled cavities, the typical
dissipation channels include the spontaneous emission of atoms
 and the cavity decay. When these decoherence effects
are taken into account and under the assumptions that the decay channels are independent, the master equation of the whole system
can be expressed by the Lindblad form \cite{scully}
\begin{eqnarray}\label{gamma}
\dot{\rho}&=&-i[H_{I},\rho]-\sum_{i=1}^3\frac{\kappa}{2}(a_i^\dag a_i\rho-2a_i\rho
a_i^\dag+\rho a_i^\dag
a_i)\nonumber\\&&-\sum_{j=0,1}\sum_{n=1}^3\frac{\gamma_n^{ej}}{2}(\sigma_{ee}^n\rho-2\sigma_{je}^n\rho\sigma_{ej}^n+\rho\sigma_{ee}^n),
\end{eqnarray}
where $\kappa$ denotes the decay rate of the cavity, $\gamma_n^{ej}$
represents the branching ration of the atomic decay from level
$|e\rangle_n$ to $|j\rangle_n$ $(n=1,2,3)$ and we assume
$\gamma_n^{e0}=\gamma_n^{e1}=\gamma/2$ for
simplicity.  The relation between the average fidelities  and the decoherence parameter $\kappa/g$ under three laser with different Rabi frequencies $(0.02g,0.05g,0.1g)$ are manifested in Fig.~\ref{pw}. The solid lines relate to the resonant model and the dash-dot lines denotes the dispersive model. Comparing these two models within given parameters, we find
that the fidelity obtained from the dispersive dynamics surpasses the resonant one only
under a relatively weak driven laser (black line: 0.02g). Thus, a resonant scheme is more realistic because a relatively strong Rabi frequency leads to a short interaction time, which makes the system avoid accumulation of decoherence in time.
In the absent of decay, we can achieve the maximal fidelities 99.98\% and 99.80\% for the resonant model $(\Omega_{max}=0.05g)$ and the dispersive model $(\Omega=0.02g)$, respectively. Even for a relatively large  decay rate $\kappa=0.01g$, the average fidelity remains above $95\% $ for the resonant model and exceeds $92\% $ for the dispersive model. In recent years, considerable progress has been made in fabrication of various high-$Q$ microcavities including whispering-gallery-mode  cavities \cite{Arm,spi}, micropost
cavities and one- or two-dimensional photonic-crystal
microcavities \cite{songb,nabe}, e.g. the strong
interaction between atom and cavity has been predicted to be available in a
toroidal microcavity system with the cavity mode wavelength of about 852 nm
\cite{spi}, where the quality factors in excess of $10^8$ can be obtained corresponding to parameters $g\sim2\pi\times
750$ MHz, $\gamma \sim 2\pi\times2.62$ MHz, $\kappa\sim 2\pi\times3.5$ MHz.
In Ref. \cite{Not}, large-scale arrays of ultrahigh-$Q$ coupled nanocavities are also achieved
with the parameters ($g,\gamma,\kappa)\sim(2.5\times10^9,1.6\times10^7,4\times10^5$) Hz. By substituting these parameters into the Eqs.~(\ref{He}) and (\ref{gamma}), we obtain the average fidelity of the Fredkin gate are 98.03\% and 97.98\% for the resonant case and 96.53\% and 98.06\% for the dispersive case, respectively. These values may not satisfy the condition for tolerating error threshold of about $10^{-5}$ \cite{pre}, but they are relatively high in the sense of multi-qubit quantum information processing.

\section{Summary}
In summary, we have presented two efficient proposals for one-step implementation of the genuine Fredkin gate via trapping three identical atoms in a coupled-cavity array. The schemes need neither local operations nor ancillary levels and they are easily controlled and robust against the typical decoherence sources in cavity QED system. It is worth noting again that the dynamical characteristic discussed in the current scheme is particular in the coupled three-cavity array system, which is incapable to be simply  duplicated in a single cavity.
 Our work may be useful for
the quantum information processing in the near future.

\begin{center}{  {ACKNOWLEDGMENT}}
\end{center}
This work is supported by Fundamental Research Funds for the Central Universities under Grant Nos. 11QNJJ009 and 12SSXM001,  National
Natural Science Foundation of China under Grant Nos. 11204028, 11175044, and 11074079, and National Research
Foundation and Ministry of Education, Singapore (Grant
No. WBS: R-710-000-008-271). X. Q. Shao is also supported in part by the Government of China through CSC.

\end{document}